\documentclass[12pt]{iopart}
\usepackage{graphics}
\usepackage{epstopdf}

\input{Qcircuit}

\newtheorem{definition}{Definition}[section]
\begin{document}

\title{New Designs of Universal Reversible Gate Library}

\author{Rasha Montaser$^{a1,2}$, Ahmed Youne$^{b1,3}$ and Mahmoud Abdel-Aty$^{c2,4}$ }

\address{$^1$ Department of Mathematics and Computer Science,
Faculty of Science, Alexandria University, Egypt}
\address{$^2$ Zewail City of Science and
Technology, University of Science and Technology, Cairo, Egypt}
\address{$^3$ School
of Computer Science, University of Birmingham, Birmingham, B15
2TT, United Kingdom}
\address{$^4$ Department of Mathematics, Faculty of Science, Sohag
University, Sohag, Egypt}
\ead{$^a$rashamontaser@gmail.com, $^b$ayounes@alexu.edu.eg and $^c$mabdelaty@zewailcity.edu.eg}
\vspace{10pt}
%\begin{indented}
%\item[]Dec. 2015
%\end{indented}

\begin{abstract}
We present new algorithms to synthesize exact universal
reversible gate library for various types of gates and costs. We
use the powerful algebraic software GAP for implementation and
examination of our algorithms and the reversible logic synthesis
problems have been reduced to group theory problems. It is shown
that minimization of arbitrary cost functions of gates and orders
of magnitude are faster than its previously counterparts for
reversible logic synthesis. Experimental results show that a
significant improvement over the previously proposed synthesis
algorithm is obtained compared with the existing approaches to
reversible logic synthesis.
\end{abstract}

%\pacs{1315, 9440T}
%
 %Uncomment for keywords
\vspace{2pc}
\noindent{\it Keywords}: Reversible gates; Reversible circuits; Quantum circuits; Quantum cost; Universal library; Universal gate.
%\maketitle

%%%%%%%%%%%%%%%%%%%%%%%%%%%%%%%%%%%%%%%
\section{Introduction}
%%%%%%%%%%%%%%%%%%%%%%%%%%%%%%%%%%%%%%%
A logic circuit is synthesized using a set of elementary
components. This set is called the library of synthesis,
 where the members of this set are logic gates
 \cite{citation12}. A gate is said to be reversible if 
 it is used to synthesize a reversible function or a 
 reversible circuit \cite{citation14}. A function is
reversible if the number of inputs is equal to the number of
outputs, and each input pattern maps to a uniquely output pattern
(bijection) \cite{citation01}. Synthesizing circuits with pure
quantum gates improves the cost and the time efficiency of these
circuits since no heat dissipation and no information destroyed
from the system, in addition, we gain the advantages of quantum
computing \cite{citation14}.  The studies to improve the
efficiency of the reversible circuits have focused on either
developing algorithms that improves the quantum cost of the
circuit by optimizing the gates used to synthesize the circuits,
as in \cite{citation05,citation10,citation06}, or by defining
new universal libraries with less number of gates that leads to
circuits with better size
on average, as in \cite{citation09,citation13}.\\

Methods developed to improve the quantum cost of reversible
circuits using the $NCT$ library are proposed in many studies such
as \cite{citation10,citation17,citation19,citation20}. An
algorithm that reduces the reversible logic synthesis problem into
permutation group using group theory is presented in
\cite{citation05}. An algorithm that uses a method that finds the
tight bounds on the synthesized 3-bits reversible circuits using
$NCT$ library to reduce the quantum cost of the circuits is
presented in \cite{citation06}. A library based synthesis
methodology for reversible circuits is presented in
\cite{citation14,citation21} to introduce optimization methods by
decomposing the functions into smaller building blocks. A new gate
type that is universal for $n$-in$\slash$out reversible circuits
is proposed in \cite{citation09}. A minimal universal library
which includes only two gates, such that all the $n\times n$
reversible circuits can be synthesized by these two gates is
proposed in \cite{citation13}.\\

The aim of this paper is to propose a new universal reversible
gate library to be used in the synthesis of reversible circuits.
The results presented in this paper are implemented and tested
using the group theory algebraic software GAP \cite{citation16}.
The results shown in this paper are compared with the results
shown in \cite{citation05}, \cite{citation09}, \cite{citation21}
and \cite{citation22}. This paper is organized as follows: Sect.2
gives a short introduction to the elementary quantum gates used to
build quantum circuits and defines the terminologies used in this
paper. Sect.3 describes the proposed gate library. Sect.4 shows
the experimental results and compares the proposed universal gate
library with the libraries proposed by others. Finally the
conclusion is shown in Sect.5.\\
%%%%%%%%%%%%%%%%%%%%%%%%%%%%%%%%%%%%%%%
\section{Synthesizing Reversible Circuit}
%%%%%%%%%%%%%%%%%%%%%%%%%%%%%%%%%%%%%%%
 Here we introduce some basic definitions and terminologies used in
the synthesizing of the reversible circuit.

\begin{definition}
A Boolean function $ f$ is reversible such that,  $f:x \to y$  if
and only if each input vector $x \in X^n $ maps to a unique output
vector $y \in X^n $. For $X^n$ there are $n$ input vector
$(x_1,x_2,x_3,\ldots,x_n)$ and $n$ output vector
$(y_1,y_2,y_3,\ldots,y_n)$. For $n$- inputs and $n$-outputs, there
are $(2^n)!$ reversible functions, $\forall X  = \{ 0,1\}  $ and $
n \in  Z$ \cite{citation06}.
\end{definition}

\begin{definition}
\({C}^{n } \)\({NOT} \)  is the main reversible gate that is
commonly used to synthesize any reversible circuit. It is defined
as
  follows,
\begin{eqnarray}
\begin{array}{l}
C^nNOT (x_1,x_2, \ldots, x_{n-1};f_{in}) =
 C^nNOT (y_1,y_2, \ldots, y_{n-1};f_{out}),
 \end{array}
\label{eqn1}
\end{eqnarray}
where $y_i=x_i$ for 1$ \le i \le n-1$ and $f_{out}=f_{in}\oplus
x_1x_2 \ldots x_{n-1}$ $ \forall n \in Z$ , $x \in X^n$ and
$X=\{0,1\}$. $x_1x_2 \ldots x_{n-1}$ are called the control bits
and $f_{in}$ is called the target bit \cite{citation06}.
\end{definition}

\begin{definition}
The minimum cost $Minc(g)$   is the realization of reversible gate
$g$ with a minimum cost  $Minc(g)$, such that; there is no
realization with cost less than the $Minc(g)$
\cite{citation05}.
\end{definition}

\begin{definition}
The cost of the circuit is the summation of the total costs of all
the gates used to synthesize that circuit
\cite{citation05}.
\end{definition}

\begin{definition}
A  gate $g$ is said to be reversible if it computes a reversible
function $f$  and is bijective
\cite{citation06,citation08}.
\end{definition}

\begin{definition}
A logic gate $g$  is said to be universal if it is sufficient to
synthesize an arbitrary logical operation on these $n$ inputs
\cite{citation12}.
\end{definition}

\begin{definition}
The set of reversible gates that can be used to build a reversible
circuit is called gate library $ L$
\cite{citation06}.
\end{definition}

\begin{definition}
The universal library of synthesis is the smallest set of building
blocks used in the synthesis process
\cite{citation12}.
\end{definition}

\begin{definition}
A universal reversible gate library $L$ is a set of reversible
gates such that any reversible $n$-bit circuit can be synthesized
by cascading gates in $L$.
\cite{citation09}.
\end{definition}

\begin{definition}
A universal reversible gate sub library $SL \subseteq  L$ is a set
of reversible gates such that any $n$-bit reversible circuit can
be synthesized using $SL$.
\cite{citation09}.
\end{definition}

\begin{definition}
Utilization is the percentage of universal sub libraries from all sub libraries $SL$
 of a certain library.\cite{citation09}.
\end{definition}

\begin{definition}
A permutation  \(\sigma\) is a bijection such that,    $\sigma
{\rm :}A \to A$   which maps an input to an output from a finite
set \({\rm  = \{ 1,2,} \ldots {\rm ,}N{\rm \} }\), which can be
written as follows,

\begin{eqnarray}
\begin{array}{l}
\sigma {\rm  = }\left( {\begin{array}{*{20}c}
   1  \\
   {\sigma (1)}  \\
\end{array}\begin{array}{*{20}c}
   2  \\
   {\sigma (2)}  \\
\end{array}\begin{array}{*{20}c}
   3  \\
   {\sigma (3)}  \\
\end{array}\begin{array}{*{20}c}
    \ldots   \\
    \ldots   \\
\end{array}\begin{array}{*{20}c}
   N  \\
   {\sigma (N)}  \\
\end{array}} \right)
 \end{array}
\label{eqn3}
\end{eqnarray}
The top row can be eliminated and written as follows,
\begin{eqnarray}
\begin{array}{l}
\sigma {\rm  = }\left( {\begin{array}{*{20}c}
   {\sigma (1)} & {\sigma (2)} & {\sigma (3)} & {\begin{array}{*{20}c}
    \ldots  & {\sigma (N)}  \\
\end{array}}  \\
\end{array}} \right)
 \end{array}
\label{eqn4}
\end{eqnarray}
Another notation having a permutation in the form of $\left( {\begin{array}{*{20}c}
   1  \\
   8  \\
\end{array}\begin{array}{*{20}c}
   2  \\
   2  \\
\end{array}\begin{array}{*{20}c}
   3  \\
   1  \\
\end{array}\begin{array}{*{20}c}
   4  \\
   4  \\
\end{array}\begin{array}{*{20}c}
   5  \\
   7  \\
\end{array}\begin{array}{*{20}c}
   6  \\
   5  \\
\end{array}\begin{array}{*{20}c}
   7  \\
   6  \\
\end{array}\begin{array}{*{20}c}
   8  \\
   3  \\
\end{array}} \right)$, it can be written as (1,8,3)(5,7,6), this notation is called the product of disjoint cycles  \cite{citation06}. 
\end{definition}
%%%%%%%%%%%%%%%%%%%%%%%%%%%%%%%%%%%%%%%
\subsection{Reversible Gates}
%%%%%%%%%%%%%%%%%%%%%%%%%%%%%%%%%%%%%%%
The reversible gates that can be used to synthesize any 3-bit
reversible circuits are: NOT ($N$) gate, Feynman ($C$) gate,
Toffoli ($T^3$) gate, Fredkin ($F$) gate, Peres ($P$) gate, the
$G^3$ gate and the square-root NOT gate which are the
controlled-$v$ ($v$) and the controlled-$v^\dag$ ($u$), such that
$v.v=u.u=N$, $u.v=v.u=I$, $v.N=N.v=u$, $u.N=N.u=v$ and $I$ is the
identity gate \cite{citation05,citation09}. For a 3-bit
reversible circuit, there are three different $N$ gates, six
different $C$ gates as shown in
Fig.\ref{fig1}, three different $T^3$ gates, three different $F$ gates as shown in
Fig.\ref{fig1-2}, six different $P$ gates
as shown in  Fig.\ref{fig2}, six different $G^3$ gates as shown in
Fig.\ref{fig2-2}, six different $v$ gates and six different $u$
gates as shown in  Fig.\ref{fig3}  \cite{citation09}.\\

\begin{figure}
\begin{center}
\[
\Qcircuit @C=1.5em @R=0.5em @!R{
\lstick{x_1}   &\targ      &\qw            &\qw            &\ctrl{1}       &\ctrl{2}       &\qw            &\targ      &\qw  &\targ           &\qw &\rstick{y_1}\\
\lstick{x_2}   &\qw            &\targ      &\qw            &\targ      &\qw            &\ctrl{1}       &\ctrl{-1}      &\targ      &\qw                    &\qw &\rstick{y_2}\\
\lstick{x_3}   &\qw            &\qw            &\targ      &\qw            &\targ      &\targ      &\qw            &\ctrl{-1}      &\ctrl{-2}      &\qw &\rstick{y_3}\\
&N_1^3 &N_2^3&N_3^3&C_{1,2}^3 &C_{1,3}^3 &C_{2,3}^3 &C_{2,1}^3 &C_{3,2}^3 &C_{3,1}^3 
}
\]
\caption{The possible $N^3$ and $C^3$  gates for 3-bits reversible circuit.} \label{fig1}
\end{center}
\end{figure}
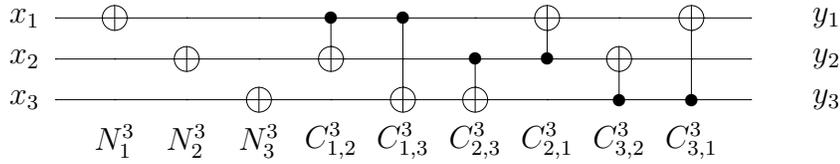

\begin{figure}
\begin{center}
\[
\Qcircuit @C=1.5em @R=0.5em @!R{
\lstick{x_1}         &\ctrl{2}       &\ctrl{1}       &\targ      &\qw  &\ctrl{1}   &\qw    &\qswap\qwx[1]  &\qw &\qswap\qwx[1] &\qw              &\rstick{y_1}\\
\lstick{x_2}   &\ctrl{1}       &\targ      &\ctrl{-1}      &\qw     &\qswap\qwx[1]&\qw  &\ctrl{1}   &\qw    &\qswap\qwx[1]  &\qw    &\rstick{y_2}\\
\lstick{x_3}   &\targ      &\ctrl{-1}      &\ctrl{-2}      &\qw       &\qswap &\qw    &\qswap &\qw    &\ctrl{0}&\qw                &\rstick{y_3}\\
&T_{1,2,3}^3 &\   \ \   T_{1,3,2}^3 &\   \ \ \  T_{2,3,1}^3&&F_{1,2,3}^3 &&F_{2,1,3}^3 &&F_{3,2,1}^3 &
}
\]
\caption{The possible $T^3$ and $F^3$  gates for 3-bits reversible circuit.}
\label{fig1-2}
\end{center}
\end{figure}
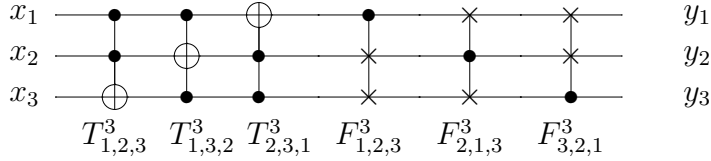

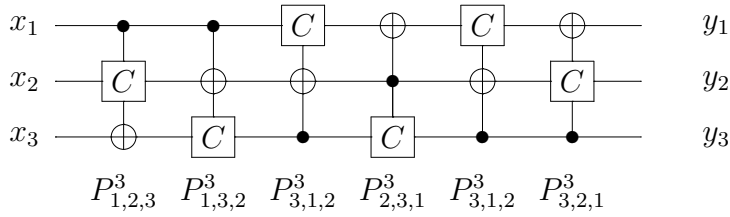
\begin{figure}
\begin{center}
\[
\Qcircuit @C=1.5em @R=0.5em @!R{
\lstick{x_1}       &\ctrl{1}           &\ctrl{1}           &\gate{C}\qwx[1]            &\targ\qwx[1]           &\gate{C}\qwx[1]        &\targ\qwx[1]       &\qw        &\rstick{y_1}\\
\lstick{x_2}     &\gate{C}\qwx[1]             &\targ  \qwx[1]     &\targ\qwx[1]       &\ctrl{1}\qwx[1]    &\targ\qwx[1]               &\gate{C}\qwx[1]                    &\qw        &\rstick{y_2}\\
\lstick{x_3}      &\targ       &\gate{C}               &\ctrl{0}           &\gate{C}           &\ctrl{0}       &\ctrl{0}               &\qw        &\rstick{y_3}\\
&P_{1,2,3}^3 & P_{1,3,2}^3&P_{3,1,2}^3&P_{2,3,1}^3&P_{3,1,2}^3&P_{3,2,1}^3
}
\]
\caption{The possible $P^3$ gates for 3-bits reversible circuit.}
\label{fig2}
\end{center}
\end{figure}

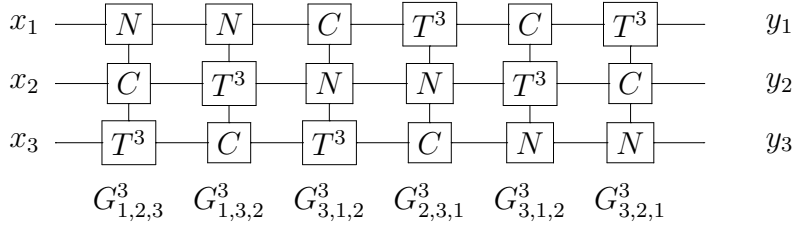
\begin{figure}
\begin{center}
\[
\Qcircuit @C=1.5em @R=0.5em @!R{
\lstick{x_1}        &\gate{N}\qwx[1]            &\gate{N}\qwx[1]            &\gate{C}\qwx[1]            &\gate{T^3}\qwx[1]          &\gate{C}\qwx[1]        &\gate{T^3}\qwx[1]      &\qw        &\rstick{y_1}\\
\lstick{x_2}      &\gate{C}\qwx[1]          &\gate{T^3}\qwx[1]      &\gate{N}\qwx[1]        &\gate{N}\qwx[1]    &\gate{T^3}\qwx[1]              &\gate{C}\qwx[1]                    &\qw        &\rstick{y_2}\\
\lstick{x_3}         &\gate{T^3}            &\gate{C}               &\gate{T^3}     &\gate{C}           &\gate{N}       &\gate{N}               &\qw        &\rstick{y_3}\\
&G_{1,2,3}^3 & G_{1,3,2}^3&G_{3,1,2}^3&G_{2,3,1}^3&G_{3,1,2}^3&G_{3,2,1}^3
}
\]
\caption{The possible $G^3$  gates for 3-bits reversible circuit.}
\label{fig2-2}
\end{center}
\end{figure}

\begin{figure}
\begin{center}
\[
\Qcircuit @C=0.7em @R=0.5em @!R{
\lstick{x_1}   &\ctrl{1}           &\ctrl{1}           &\qw                &\gate{v}\qwx[1]    &\qw                &\gate{v}\qwx[1]    &\ctrl{1}           &\ctrl{1}           &\qw                &\gate{u}\qwx[1]    &\qw                &\gate{u}\qwx[1]    &\qw        &\rstick{y_1}\\
\lstick{x_2}   &\gate{v}           &\qw\qwx[1]     &\ctrl{1}           &\ctrl{0}           &\gate{v}\qwx[1]    &\qw\qwx[1]     &\gate{u}       &\qw\qwx[1]     &\ctrl{1}           &\ctrl{0}           &\gate{u}\qwx[1]    &\qw\qwx[1]     &\qw        &\rstick{y_2}\\
\lstick{x_3}   &\qw                &\gate{v}           &\gate{v}           &\qw                &\ctrl{0}           &\ctrl{0}           &\qw                &\gate{u}       &\gate{u}       &\qw                &\ctrl{0}           &\ctrl{0}           &\qw        &\rstick{y_3}\\
&v_{1,2} &v_{1,3} &v_{2,3} &v_{2,1}& v_{3,2}&v_{3,1}&u_{1,2} &u_{1,3} &u_{2,3} &u_{2,1}& u_{3,2}&u_{3,1}
}
\]
\caption{The possible $v$ and $u$ gates for 3-bits reversible circuit.}
\label{fig3}
\end{center}
\end{figure}
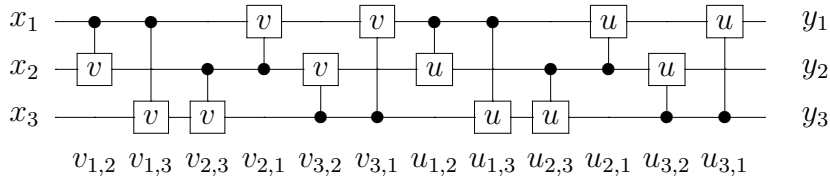

The $N$ gate is used to flip the input bits unconditionally with
quantum cost equal to zero \cite{citation21}.  Eqn.\ref{eqn5}
shows the functionality of the $N$ gate. There are 8 circuits that
can be realized using the three possible $N$ gates
\cite{citation09}.
\begin{eqnarray}
\begin{array}{l}
N_j^3 : y_j=x_j\oplus1, y_k=x_k, y_l=x_l,\\
\\
 N_1^3: (x_1 ,x_2 ,x_3 )  {\rm }{\rm } {\rm \to (1,5)(2,6)(3,7)(4,8)}, \\
 N_2^3: {\rm (}x_1 ,x_2 ,x_{\rm 3} {\rm )} {\rm }{\rm } {\rm \to (1,3)(2,4)(5,7)(6,8)},\\
 N_3^3: {\rm (}x_1 ,x_2 ,x_3 {\rm )} {\rm }{\rm } {\rm \to (1,2)(3,4)(5,6)(7,8)}, \\
 \end{array}
\label{eqn5}
\end{eqnarray}
where $ j ,k$ and $ l \in \{1,2,3\}$  in any order.\\

The $C$ gate is used to flip the target bit if the controlled bit
is set to 1 with quantum cost equals to 1 \cite{citation21}.
Eqn.\ref{eqn6} shows the functionality of the $C$ gate. There are
168 circuits that can be realized by the six possible $C$ gates
\cite{citation09}.
\begin{eqnarray}
\begin{array}{l}
C^3_{j,k}: y_j=x_j, y_k=x_k \oplus x_j, y_l=x_l,\\
\\
{C}_{{ 1,2}}^{ 3} { :(x}_{ 1} { ,x}_{ 2} { ,x}_{ 3} { )} \to {  (5,7)(6,8)}, \\
{ C}_{{ 1,3}}^{ 3} { :(x}_{ 1} { ,x}{}_{ 2}{ ,x}_{ 3} { )} \to { (5,6)(7,8)}, \\
{ C}_{{ 2,3}}^{ 3} { :(x}_{ 1} { ,x}_{ 2} { ,x}_{ 3} { )} \to { (3,4)(7,8)}, \\
{ C}_{{ 2,1}}^{ 3} { :(x}_{ 1} { ,x}_{ 2} { ,x}_{ 3} { )} \to { (3,7)(4,8)},\\
{ C}_{{ 3,2}}^{ 3} { :(x}_{ 1} { ,x}_{ 2} { ,x}_{ 3} { )} \to {  (2,4)(6,8)}, \\
{ C}_{{ 3,1}}^{ 3} { :(x}_{ 1} { ,x}_{ 2} { ,x}_{ 3} { )} \to { (2,6)(4,8)}.\\
\end{array}
\label{eqn6}
\end{eqnarray}
The $T^3$ gate is used to flip the target bit if the control bits
are set to 1 with quantum cost equals to 5 \cite{citation21}.
Eqn.\ref{eqn7} shows the functionality of the $T^3$ gate. There
are 24 circuits that can be realized by the three possible $T^3$
gates \cite{citation09}.
\begin{eqnarray}
\begin{array}{l}
T^3_{j,k,l}: y_j=x_j, y_k=x_k, y_l=x_l \oplus x_jx_k,\\
\\
 T_{1,2,3}^3 {\rm :(}x_1 ,x_2 ,x_3 {\rm )} \to { (7,8)}, \\
 T_{1,3,2}^3 {\rm :(}x_1 ,x_2 ,x_3 {\rm )} \to { (6,8)}, \\
 T_{3,2,1}^3 {\rm :(}x_1 ,x_2 ,x_3 {\rm )} \to { (4,8)}. \\
 \end{array}
\label{eqn7}
\end{eqnarray}
The $F$ gate is used to perform conditional swap on two of its
inputs if the third input is set to 1 with quantum cost equals to
5 \cite{citation20}. Eqn.\ref{eqn8} shows the functionality of the
$F$ gate. There are 6 circuits that can be realized by the three
possible $F$ gates \cite{citation09}.
\begin{eqnarray}
\begin{array}{l}
F^3_{j,k,l}: y_j=x_j,\\
\ \ \ \ \ \ \ \ \ x_j  = \left\{ {\begin{array}{*{20}c}
   {\begin{array}{*{20}c}
   1 & {y_k  = x_l ,y_l  = x_k }  \\
\end{array}}  \\
   {\begin{array}{*{20}c}
   0 & {y_k  = x_k ,y_l  = x_l }  \\
\end{array}}  \\
\end{array}} \right.

\\
\\
F_{1,2,3}^3 {\rm :(}x_1 ,x_2 ,x_3 {\rm )} \to { (6,7)}, \\
F_{2,1,3}^3 {\rm :(}x_1 ,x_2 ,x_3 {\rm )} \to { (4,7)}, \\
F_{3,2,1}^3 {\rm :(}x_1 ,x_2 ,x_3 {\rm )} \to { (4,6)}. \\
 \end{array}
\label{eqn8}
\end{eqnarray}
The $P$ gate combines the functions of $T^3$ gate and $C$ gate in
a single gate; it acts on an arbitrary 3-bits $x_j$, $x_k$ and
$x_l$, $C$ gate is applied on $x_j$ and $x_k$ using $x_j$ as a
controller bit and $x_k$ as a target bit, then $T^3$ gate is
applied on $x_j$, $x_k$ and $x_l$ using $x_j$ and $x_k$ as
controller bits and $x_l$ as a target bit. The quantum cost of $P$
is 4 \cite{citation20}. Eqn.\ref{eqn9} shows the functionality of
the $P$ gate. There are 5040 circuits that can be realized by the
six possible $P$ gates \cite{citation09}.
\begin{eqnarray}
\begin{array}{l}
P^3_{j,k,l}: y_j=x_j, y_k=x_j \oplus x_k,y_l=x_l \oplus x_jx_k,\\
\\
 P_{123 } : (x_1 , x_2 , x_3 )  \to  (5, 7, 6, 8), \\
 P_{132}  : (x_1 , x_2 , x_3 )  \to  (5, 6, 7, 8), \\
 P_{213}  : (x_1 , x_2 , x_3 )  \to   (3, 7, 4, 8), \\
 P_{231}  : (x_1 , x_2 , x_3 )  \to  (3, 4, 7, 8), \\
 P_{312}  : (x_1 , x_2 , x_3 )  \to  (2, 6, 4, 8), \\
 P_{321}  : (x_1 , x_2 , x_3 )  \to  (2, 4, 6, 8). \\
 \end{array}
\label{eqn9}
\end{eqnarray}
The $G^3$ gate combines the function of $N$, $C$ and $T^3$ gates
in a single gate; it acts on an arbitrary 3-bits $x_j$, $x_k$ and
$x_l$. $x_l$ is flipped if $x_j$ and $x_k$ are set to 1, then
$x_k$ is flipped if $x_j$ is set to 1, finally the bit $x_j$ is
flipped unconditionally. The quantum cost of $G^3$ is 5
\cite{citation09}. Eqn.\ref{eqn9-2} shows the functionality of the
$G^3$ gate. There are 40320 circuits that can be realized by the
six possible $G^3$ gates, thus $G$ gate is universal
\cite{citation09}.
\begin{eqnarray}
\begin{array}{l}
G^3_{j,k,l}: y_j=x_j \oplus 1, y_k=x_j \oplus x_k,y_l=x_l \oplus x_jx_k,\\
\\
G_{123 } : (x_1 , x_2 , x_3 )  \to  (1,5,3,7,2,6,4,8), \\
G_{132}  : (x_1 , x_2 , x_3 )  \to  (1,5,2,6,3,7,4,8), \\
 G_{213}  : (x_1 , x_2 , x_3 )  \to   (1,3,5,7,2,4,6,8), \\
 G_{231}  : (x_1 , x_2 , x_3 )  \to  (1,3,2,4,5,7,6,8), \\
 G_{312}  : (x_1 , x_2 , x_3 )  \to  (1,2,5,6,3,4,7,8), \\
 G_{321}  : (x_1 , x_2 , x_3 )  \to  (1,2,3,4,5,6,7,8). \\
 \end{array}
\label{eqn9-2}
\end{eqnarray}
%%%%%%%%%%%%%%%%%%%%%%%%%%%%%%%%%%%%%%%%%%%%
\subsection{Universal Libraries}

Many universal libraries have been defined from the combination of the reversible gates, such as $NCT$ (NOT-
 Feynman- Toffoli), $NCTF$ (NOT- Feynman- Toffoli- Fredkin),  $NCP$ (NOT- Feynman- Peres),  $NCPF$ (NOT- 
Feynman- Peres- Fredkin), $NCPT$  (NOT- Feynman- Peres-
 Toffoli), $N$$ T$ (NOT- Toffoli), $NP$ (NOT- Peres), $NCF$
 (N-OT- Feynman- Fredkin), $NFT$ (NOT-  Feynman-  Toffoli), $NCTPF$ (NOT-  Feynman- Toffoli- Peres- Fredkin) and $G$-gate Library \cite{citation05,citation09,citation22}.
%%%%%%%%%%%%%%%%%%%%%%%%%%%%%%%%%%%%%%

\section{The Proposed Universal Reversible Library}
\label{sec:2}
%%%%%%%%%%%  Section2 %%%%%%%%%%%%%%%%%%%%%

This section proposes a new universal $n$-bit reversible gate
$R^n$ for $n$-bits input/output reversible circuits, which is a 
universal gate on its own. To reduce the quantum cost of the
circuits synthesized with $R^n$ gate, $N$ gate might be added
to form another library called $NR^n$ which is also universal.\\

%%%%%%%%%%%%%%%%%%%%%%%%%%%%%%
\subsection{1-bit Gate ($R^1$ gate)}
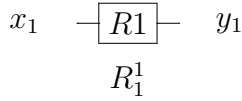
\begin{figure}
\begin{center}
\[
\Qcircuit @C=0.7em @R=0.5em @!R{
\lstick{x_1}&   &\gate{R1}      &\qw     &\rstick{y_1}\\
&& R_{1}^1}
\]
\caption{The one possible $R^1 $ gate for 1-bit reversible circuit.}
\label{fig6}
\end{center}
\end{figure}

$R^1$ is a 1-bit gate. It acts as $N$ gate which inverses the
input bit unconditionally. There are 2 possible 1-bit input/output
reversible circuits. The $R^1$ gate is sufficient to realize these
two circuits. For 1-bit reversible circuits built using $R$-gate
library, there is one $R^1$ gate as shown in Fig.\ref{fig6},  it
function as shown in Eqn.\ref{eqn10} and its quantum cost is 0. \\

\begin{eqnarray}
\begin{array}{l}
R_1^1 :(x_1 )  =  x_1  \oplus 1=(1,2).\\
 \end{array}
\label{eqn10}
\end{eqnarray}

%%%%%%%%%%%%%%%%%%%%%%%%

\subsection{2-bits Gate ($R^2$ gate)}
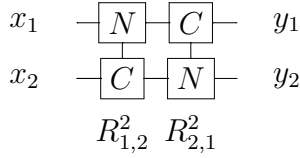
\begin{figure}
\begin{center}
\[
\Qcircuit @C=0.7em @R=0.5em @!R{
\lstick{x_1}& &\gate{N}\qwx[1]  &\gate{C}\qwx[1] &\qw  &\rstick{y_1}\\
\lstick{x_2}& &\gate{C}\qw  &\gate{N}\qw &\qw  &\rstick{y_2}\\
&&R_{1,2}^2 &R_{2,1}^2
}
\]
\caption{The two possible $R^2 $ gates for 2-bits reversible circuit.}
\label{fig7}
\end{center}
\end{figure}

$R^2$ is a 2-bits gate. It acts as a combination between $N$ gate
and $C$ gate. It uses one bit as a controller to flip the other
bit and then it flips the controller bits unconditionally. For 2
input/output reversible circuits there are 24 possible circuits.
$R^2$ gate is sufficient to realize these 24 circuits. For 2-bits
reversible circuits  built using $R$-gate library, there are two
possible $R^2$ gates as shown in  Fig.\ref{fig7}, they function
as shown in Eqn.\ref{eqn11} and their quantum cost is 1. 
\begin{eqnarray}
\begin{array}{l}
R_{j,k}^2: y_j=x_j \oplus 1 ,\\
\ \ \ \ \ \ \ \ y_k=x_k \oplus x_j,\\
\\
R_{1,2}^2  :(x_1 ,x_2 )   \to  (1, 3 ,2, 4), \\
R_{2,1}^2 :(x_1 ,x_2 ) \to  (1, 2, 3 ,4), \\
 \end{array}
\label{eqn11}
\end{eqnarray}
where $ j$  and  $k$  $ \in \{1,2\}$  in any order.
%%%%%%%%%%%%%%%%%%%%%%%%%%%%%%%%%%
\subsection{3-bits Gate ($R^3$ gate)}
$R^3$ is a 3-bits gate. It combines the action of the three gates
$N$, $C$ and $T^3$. It acts on an arbitrary 3-bits $x_j$, $x_k$
and $x_l$ in any order. First $x_j$ and $x_k$ are used as
controller bits to flip $x_l$, then $x_j$ is used as a controller
bit to flip $x_k$ , after that $x_l$ is flipped unconditionally
and finally $x_l$ is used as a controller bit to flip $x_j$. For
3-bits reversible circuits built using $R$-gate library, there are
six possible $R^3$ gates as shown in  Fig.\ref{fig8} and they
function as shown in Eqn.\ref{eqn12}. \\
\begin{figure}
\begin{center}
\[
\Qcircuit @C=0.5em @R=0.5em @!R{
\lstick{x_1} &\gate{C_{2,1}}\qwx[1]                &\qw        &\gate{C_{3,1}}\qwx[1]       &\qw       &\gate{C_{2,1}}\qwx[1]          &\qw        &\gate{C_{3,1}}\qwx[1]       &\qw       &\gate{T^3N}\qwx[1]             &\qw        &\gate{T^3N}\qwx[1]                 &\qw        &\rstick{y_1}\\
\lstick{x_2} &\gate{T^3N}\qwx[1]               &\qw        &\gate{T^3N}\qwx[1]                 &\qw        &\gate{C_{3,2}}\qwx[1]              &\qw        &\gate{C_{1,2}}\qwx[1]              &\qw        &\gate{C_{3,2}}\qwx[1]              &\qw        &\gate{C_{1,2}}\qwx[1]      &\qw        &\rstick{y_2}\\
\lstick{x_3} &\gate{C_{1,3}}               &\qw &\gate{C_{2,3}}                &\qw        &\gate{T^3N}            &\qw        &\gate{T^3N}                &\qw        &\gate{C_{1,3}}                 &\qw        &\gate{C_{2,3}}                 &\qw        &\rstick{y_3}\\
& R_{1,2,3}^3 &&  R_{3,2,1}^3&&  R_{3,1,2}^3&&  R_{1,3,2}^3&&  R_{2,3,1}^3&&  R_{2,1,3}^3
}
\]
\caption{The six possible $R^ 3 $ gates for 3-bits reversible circuit.}
\label{fig8}
\end{center}
\end{figure}
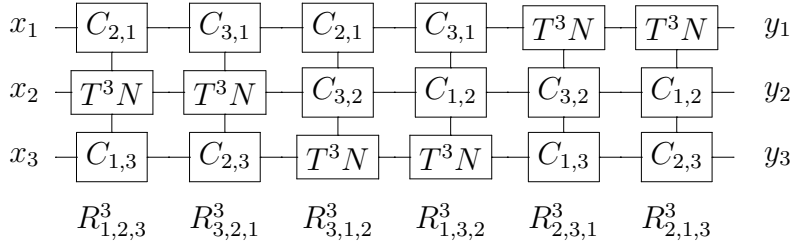
\begin{eqnarray}
\begin{array}{l}
R_{j,k,l}^3: y_j=x_j \oplus x_k \oplus x_j . x_l\oplus 1 ,\\
\ \ \ \ \ \ \ \ \   y_k=x_k \oplus x_j . x_l\oplus 1, \\
\ \ \ \ \ \ \ \ \   y_l=x_l \oplus x_j,\\
\\
 R_{1,2,3}^3  :(x_1 ,x_2 ,x_3 ) \to (1 ,7 ,6 , 5 ,4 ,2 ,8 ,3), \\
 R_{3,2,1}^3 :(x_1 ,x_2 ,x_3 ) \to (1 ,4 ,6 ,2 ,7 ,5 ,8 ,3), \\
 R_{3,1,2}^3 :(x_1 ,x_2 ,x_3 ) \to (1 ,4 ,7 ,3 ,6 ,5 ,8 ,2),   \\
 R_{1,3,2}^3 :(x_1 ,x_2 ,x_3 ) \to (1 ,6 ,7 ,5 ,4 ,3 ,8 ,2), \\
 R_{2,3,1}^3 :(x_1 ,x_2 ,x_3 ) \to (1 ,6 ,4 ,2 ,7 ,3 ,8 ,5), \\
 R_{2,1,3}^3 :(x_1 ,x_2 ,x_3 ) \to (1,7 ,4 ,3 ,6 ,2 ,8 ,5), \\
 \end{array}
\label{eqn12}
\end{eqnarray}
where $ j ,k$ and $ l \in \{1,2,3\}$  in any order.\\

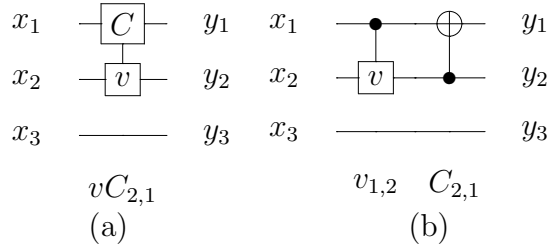
\begin{figure} [htbp]
\centerline{
\begin{tabular}{p{3cm}p{3cm}}
\Qcircuit @C=0.7em @R=0.5em @!R{
\lstick{x_1}&   &\gate{C}\qwx[1]    &\qw            &\rstick{y_1}\\
\lstick{x_2}&   &\gate{v}   &\qw    &\rstick{y_2}\\
\lstick{x_3}&    &\qw   &\qw        &\rstick{y_3}\\
&&vC_{2,1}
}&
\Qcircuit @C=0.7em @R=0.7em @!R{
\lstick{x_1}&   &\ctrl{1}   &\qw    &\targ&\qw&\rstick{y_1}\\
\lstick{x_2}&   &\gate{v}       &\qw&\ctrl{-1}  &\qw&\rstick{y_2}\\
\lstick{x_3}&   &\qw    &\qw&\qw    &\qw                &\rstick{y_3}\\
&& v_{1,2} &&\ C_{2,1}
}\\ \ \ \ (a)&\ \ \ \ \ \ \ \ \ (b)
\end{tabular}
} %100 percent
\caption{The circuit representation for the decomposition of the
$vC_{2,1}$ gate, where: (a) the representation of the gate, and (b)
the decomposition of the gate into it’s two components
$v_{1,2}$ gate and $C_{2,1}$ gate.} \label{fig9-1}
\end{figure}
\begin{figure} [htbp]
\centerline{
\begin{tabular}{p{2.5cm}p{3.7cm}p{2.8cm}}
\Qcircuit @C=0.35em @R=0.5em @!R{
\lstick{x_1}& &\gate{C_{2,1}}\qwx[1]  &\qw &\rstick{y_1}\\
\lstick{x_2}& &\gate{T^3N}\qwx[1]  &\qw &\rstick{y_2}\\
\lstick{x_3}& &\gate{C_{1,3}}  &\qw &\rstick{y_3}\\
&&R_{1,2,3}^3
}&
\Qcircuit @C=0.35em @R=1.3em @!R{
\lstick{x_1}&   &\ctrl{1}   &\qw&\ctrl{2}       &\qw&\qw        &\qw&\targ  &\qw        &\rstick{y_1}\\
\lstick{x_2}&   &\targ\qwx[1]   &\qw&\qw    &\qw    &\targ      &\qw&\ctrl{-1}              &\qw        &\rstick{y_2}\\
\lstick{x_3}&   &\ctrl{0}       &\qw&\targ  &\qw&\qw            &\qw&\qw        &\qw        &\rstick{y_3}\\
&& T_{1,3,2}^3 &&\ \ \ \  C_{1,3} &&\ \ \ \  N_2&& \ \ \ \  \ C_{2,1}
}
&
\Qcircuit @C=.7em @R=.8em @!R{
\lstick{x_1}&   &\qw            &\ctrl{1}           &\qw        &\qw        &\gate{C}\qwx[1]            &\qw        &\rstick{y_1}\\
\lstick{x_2}&   &\gate{v}\qwx[1]    &\qw\qwx[1]         &\gate{u}   &\targ      &\gate{v}                       &\qw        &\rstick{y_2}\\
\lstick{x_3}&&\ctrl{-1} &\targ          &\ctrl{-1}          &\qw        &\qw                    &\qw        &\rstick{y_3}\\
&&v_{3,2}& \ C_{1,3}& \  u_{3,2}& \ N_2&\ \ vC_{2,1}
}\\ \\ \ \ \ \ (a)&\ \ \ \ \ \ \ \ \ \ (b)&\ \ \ \ \ \ \ \ \ \ \ \ \ \ \ \ (c)
\end{tabular}
} %100 percent
\caption{The circuit representation for the decomposition of the
$R_{1,2,3}^3$ gate where: (a) the representation of the gate, (b)
the decomposition of the gate into it’s four components two $C$
gates, one $N$ gate and one $T^3$ gate, and (c) the
optimized decomposition of $R_{1,2,3}^3$ into it’s five
elementary quantum gates.} \label{fig9}
\end{figure}
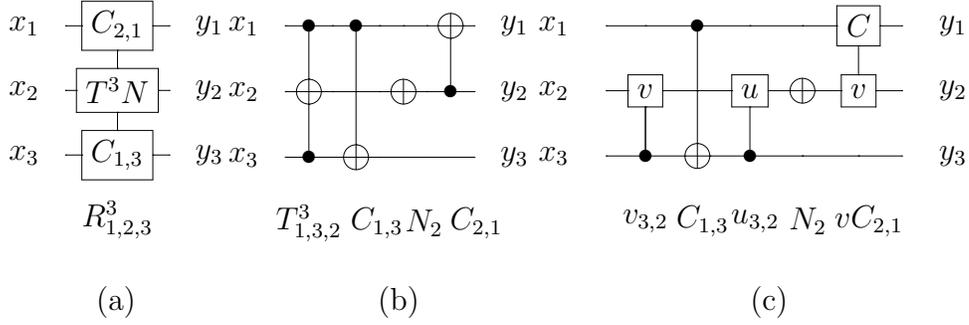
The quantum cost for the $R^3$ gate is 4, Fig.\ref{fig9} shows the
decomposition of the gate. Fig.\ref{fig9}(a) shows the gate
representation of $R^3_{1,2,3}$ gate, Fig.\ref{fig9}(b) shows the
four component gates of the $R^3_{1,2,3}$ gate, and
Fig.\ref{fig9}(c) shows the representation of the $R^3_{1,2,3}$
gate into it’s five elementary gates after applying circuit
optimization over it as defined in \cite{citation21}. The first
gate is $v_{3,2}$, the second gate is $C_{1,3}$, the third gate is
$u_{3,2}$, the fourth gate is $N_2$  and the last gate is
$vC_{2,1}$, which is a merging gate between $v_{1,2}$ and
$C_{2,1}$ in order, as shown in  Fig.\ref{fig9-1}, the process of
merging gates is defined in \cite{citation17}.\\
%%%%%%%%%%%%

%
\begin{small}
\begin{eqnarray}
\begin{array}{l}
R_{j,k,l,m}^4: \ \ \ y_j=x_j \oplus x_k \oplus x_j . x_l\oplus 1 ,\\
\ \ \ \ \ \ \ \ \  \  \ \ \  \  \  y_k=x_k \oplus x_j . x_l\oplus 1, \\
\ \ \ \ \ \ \ \ \  \  \ \ \  \  \  y_l=x_l \oplus x_j,\\
\ \ \ \ \ \ \ \ \  \  \ \ \  \  \  y_m= x_m \oplus x_j.x_k.x_l, \\
 
\\
 R_{1,2,3,4}^4 :(x_1 ,x_2 ,x_3 ,x_4 ) \to (1, 13, 11, 9,7, 3,15, 6, 2, 14, 2, 10, 8, 4, 16, 5 ), \\
 R_{3,2,1,4}^4 :(x_1 ,x_2 ,x_3 ,x_4 ) \to(1,7,11,3,13,9,15,6,2,8,12,4,14,10,16,5 ), \\
 R_{3,1,2,4}^4 :(x_1 ,x_2 ,x_3 ,x_4 ) \to (1,7,13,5,11,9,15,4,2,8,14,6,12,10,16,3), \\
 R_{1,3,2,4}^4 :(x_1 ,x_2 ,x_3 ,x_4 ) \to  (1,11,13,9,7,5,15,4,2,12,14,10,8,6,16,3 ), \\
 R_{2,3,1,4}^4 :(x_1 ,x_2 ,x_3 ,x_4 ) \to (1,11,7,3,13,5,15,10,2,12,8,4,14,6,16,9), \\
 R_{2,1,3,4}^4 :(x_1 ,x_2 ,x_3 ,x_4 ) \to(1,13,7,5,11,3,15,10,2,14,8,6,12,4,16,9 ), \\
 R_{1,2,4,3}^4 :(x_1 ,x_2 ,x_3 ,x_4 ) \to  (1,13,10,9,6,2,14,7,3,15,12,11,8,4,16,5 ), \\
 R_{3,2,4,1}^4 :(x_1 ,x_2 ,x_3 ,x_4 ) \to(1,6,10,2,13,9,14,7,3,8,12,4,15,11,16,5 ), \\

 R_{3,1,4,2}^4 :(x_1 ,x_2 ,x_3 ,x_4 ) \to(1,6,13,5,10,9,14,4,3,8,15,7,12,11,16,2), \\
 R_{1,3,2,4}^4 :(x_1 ,x_2 ,x_3 ,x_4 ) \to (1,11,13,4,2,12,14,9,7,5,15,10,8,6,16,3), \\
 R_{2,3,4,1}^4 :(x_1 ,x_2 ,x_3 ,x_4 ) \to (1,10,6,2,13,5,14,11,3,12,8,4,15,7,16,9), \\
 R_{2,1,4,3}^4:(x_1 ,x_2 ,x_3 ,x_4 )  \to  (1,13,6,5,10,2,14,11,3,15,8,7,12,4,16,9),\\
R_{1,4,2,3}^4:({x_1},{x_2},{x_3},{x_4}) \to (1,11,10,9,4,2,12,7,5,15,14,13,8,6,16,3),\\
R_{3,4,2,1}^4:({x_1},{x_2},{x_3},{x_4}) \to  (1,4,10,2,11,9,12,7,5,8,14,6,15,13,16,3),\\
R_{3,4,1,2}^4:({x_1},{x_2},{x_3},{x_4}) \to (1,4,11,3,10,9,12,6,5,8,15,7,14,13,16,2),\\
R_{1,4,3,2}^4:({x_1},{x_2},{x_3},{x_4}) \to (1,10,11,9,4,3,12,6,5,14,15,13,8,7,16,2),\\
R_{2,4,3,1}^4:({x_1},{x_2},{x_3},{x_4}) \to (1,10,4,2,11,3,12,13,5,14,8,6,15,7,16,9),\\
R_{2,4,1,3}^4:({x_1},{x_2},{x_3},{x_4}) \to  (1,11,4,3,10,2,12,13,5,15,8,7,14,6,16,9),\\
R_{4,1,2,3}^4:({x_1},{x_2},{x_3},{x_4}) \to  (1,7,6,5,4,2,8,11,9,15,14,13,12,10,16,3),\\
R_{4,3,2,1}^4:({x_1},{x_2},{x_3},{x_4}) \to  (1,4,6,2,7,5,8,11,9,12,14,10,15,13,16,3),\\
R_{4,3,1,2}^4:({x_1},{x_2},{x_3},{x_4}) \to  (1,4,7,3,6,5,8,10,9,12,15,11,14,13,16,2),\\
R_{4,1,3,2}^4:({x_1},{x_2},{x_3},{x_4}) \to  (1,6,7,5,4,3,8,10,9,14,15,13,12,11,16,2),\\
R_{4,2,3,1}^4:({x_1},{x_2},{x_3},{x_4}) \to  (1,6,4,2,7,3,8,13,9,14,12,10,15,11,16,5),\\
R_{4,2,1,3}^4:({x_1},{x_2},{x_3},{x_4}) \to  (1,7,4,3,6,2,8,13,9,15,12,11,14,10,16,5),\\
 \end{array}
\label{eqn14}
\end{eqnarray}
\end{small}

%%%%%%%%%%%%%%%%%%%%%%%%%%%%%%%%%%%%%
\subsection{4-bits Gate ($R^4$ gate)}
\begin{figure} [htbp]
\vspace*{13pt}
\centerline{
\begin{tabular}{p{2.5cm}p{2.5cm}}
\Qcircuit @C=0.35em @R=0.8em @!R{
\lstick{x_1}& &\gate{C_{2,1}}\qwx[1]  &\qw &\rstick{y_1}\\
\lstick{x_2}& &\gate{T^3N}\qwx[1] &\qw &\rstick{y_2}\\
\lstick{x_3}& &\gate{C_{1,3}}\qwx[1] &\qw &\rstick{y_3}\\
\lstick{x_4}& &\gate{T^4} &\qw &\rstick{y_4}\\
&&R_{1,2,3,4}^4
}&
\Qcircuit @C=0.5em @R=1.55em @!R{
\lstick{x_1}&&\ctrl{1}      &\qw&\qw&\ctrl{1}   &\qw        &\qw&\ctrl{2}   &\qw    &\qw        &\qw&\targ  &\qw        &\rstick{y_1}\\
\lstick{x_2}&   &\ctrl{1}   &\qw&\qw&\targ      &\qw&\qw&\qw        &\qw&\targ      &\qw&\ctrl{-1}              &\qw        &\rstick{y_2}\\
\lstick{x_3}&&\ctrl{1}      &\qw&\qw&\ctrl{-1}      &\qw&\qw&\targ  &\qw&\qw            &\qw&\qw    &\qw            &\rstick{y_3}\\
\lstick{x_4}&   &\targ      &\qw&\qw&\qw&\qw&\qw &\qw&\qw       &\qw&\qw    &\qw&\qw                &\rstick{y_4}\\
&& T_{1,2,3,4}^4 &&&\  \  \ T_{1,3,2}^3 &&&\  \  \  C_{1,3}^2 &&\  \  \ N_{2}^3&&\  \  \  C_{2,1}^2
}\\(a)&(b)
\end{tabular}
} %100 percent
\caption{The circuit representation for the decomposition of the
$R^4$ gate where: (a) the representation of the gate, (b) the
decomposition of the gate into it’s 5 components one $T^4$ gate,
one $T^3$ gate, two $C$ gates and one $N$ gate.} \label{fig10}
\end{figure}
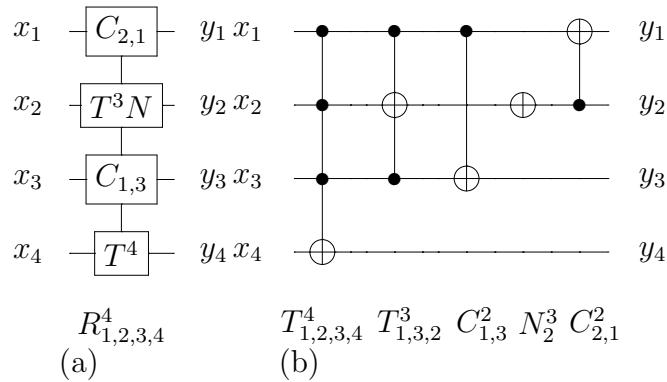

$R^4$ is 4-bits gate, which combines the action of the four gates
$N$, $C$, $T^3$ and $T^4$. The four gates are applied on an arbitary
4-bits $x_j,x_k,x_l,x_m$ in any order as follows, First the 3 bits
$x_j,x_k$ and $x_l$ are used as controller bits to flip the bit $x_m$,
then the 2 bits $x_j$ and $x_k$ are used as controller bits to
flip $x_l$. After that $x_j$ is used as a controller bit to flip
$x_k$, then $x_l$ is flipped unconditionally and finally $x_l$ is
used as a controller bit to flip $x_j$. Fig.\ref{fig10} shows the
decomposition of the gate, Fig.\ref{fig10}(a) shows the
representation of the $R_{1,2,3,4}^4$ gate, and Fig.\ref{fig10}(b)
shows the decomposition of the $R_{1,2,3,4}^4$ gate into it’s
five components. There are 24 possible $R^4$ gates. These 24 gates
are sufficient to realize the $(2^4)!$ circuits. For 4-bits
reversible circuits built using $R$-gate library, there are 24
possible $R^4$ gates, their function is shown in Eqn.\ref{eqn14} .\\

%%%%%%%%%%%%Eqn4%%%%%%%%%%%%%%%%
%
%
%%%%%%%%%%%%%%%%%%%%%%%%%%%%%%%%%%%%%%

\subsection{$n$-bits Gate ($R^n$ gate)}
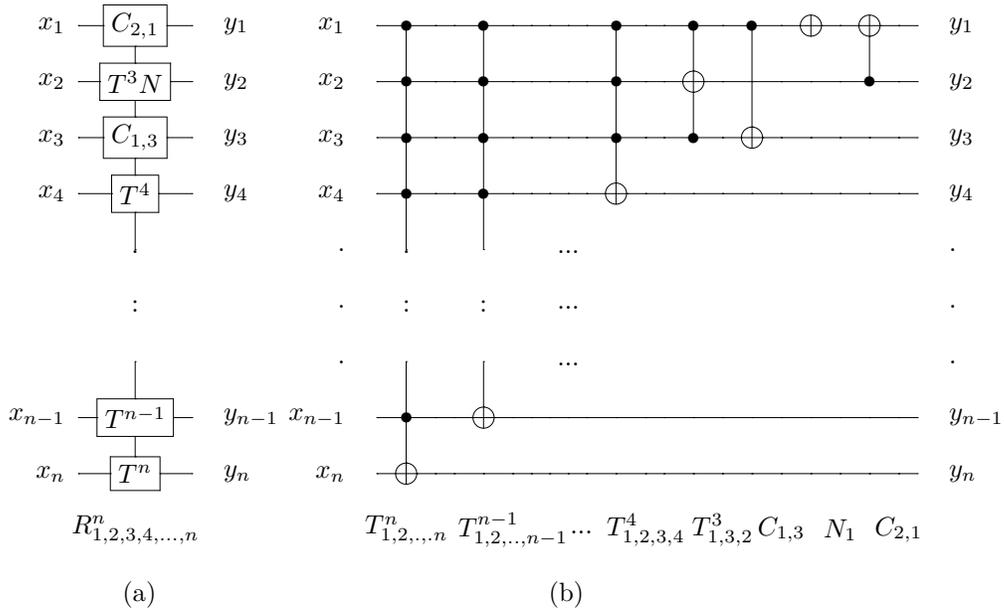
\begin{figure} [htbp]
\centerline{
\begin{footnotesize}
\begin{tabular}{p{3.3cm}p{3cm}}
\Qcircuit @C=0.7em @R=0.55em @!R{
\lstick{x_1}&\gate{C_{2,1}}\qwx[1]  &\qw &\rstick{y_1}\\
\lstick{x_2}&\gate{T^3N}\qwx[1]  &\qw &\rstick{y_2}\\
\lstick{x_3}&\gate{C_{1,3}}\qwx[1] &\qw &\rstick{y_3}\\
\lstick{x_4}&\gate{T^4}\qwx[1]  &\qw &\rstick{y_4}\\
\lstick{}&.&&&\\
\lstick{}&:&&&\\
\lstick{}&\qwx[1].&&&\\
\lstick{x_{n-1}}&\gate{T^{n-1}}\qwx[1] &\qw &\rstick{y_{n-1}}\\
\lstick{x_{n}}&\gate{T^n} &\qw &\rstick{y_{n}}\\
&R_{1,2,3,4,...,n}^n
}&
\Qcircuit @C=0.7em @R=1.3em @!R{
\lstick{x_1}&&\ctrl{1}&\qw&\qw&\ctrl{1} &\qw&\qw&\qw&\qw    &\qw&\ctrl{1}&\qw&\qw&\ctrl{1}  &\qw    &\ctrl{2}       &\qw &\targ     &\qw&\targ  &\qw    &\qw    &\rstick{y_1}\\
\lstick{x_2}&   &\ctrl{1}&\qw&\qw&\ctrl{1} &\qw&\qw&\qw&\qw&\qw &\ctrl{1}&\qw&\qw   &\targ      &\qw&\qw        &\qw    &\qw    &\qw&\ctrl{-1}          &\qw&\qw        &\rstick{y_2}\\
\lstick{x_3}&&\ctrl{1}&\qw&\qw&\ctrl{1} &\qw&\qw&\qw&\qw&\qw    &\ctrl{1}&\qw&\qw       &\ctrl{-1}      &\qw&\targ                  &\qw&\qw        &\qw&\qw        &\qw&\qw        &\rstick{y_3}\\
\lstick{x_4}&   &\ctrl{1}&\qw&\qw&\ctrl{1} &\qw&\qw&\qw&\qw&\qw &\targ  &\qw    &\qw&\qw&\qw&\qw&\qw &\qw&\qw       &\qw&\qw    &\qw            &\rstick{y_4}\\
\lstick{.}& &.&&&&&&&...&&&&&&&&&&&&&&\rstick{.}\\
\lstick{.}& &:&&&:&&&&...&&&&&&&&&&&&&&\rstick{.}\\
\lstick{.}& &\qwx[1]&&&\qwx[1]&&&&...&&&&&&&&&&&&&&\rstick{.}\\
\lstick{x_{n-1}}&&\ctrl{1}&\qw&\qw &\targ&\qw&\qw&\qw&\qw&\qw&\qw&\qw&\qw&\qw&\qw&\qw&\qw&\qw&\qw&\qw&\qw&\qw&\rstick{y_{n-1}}\\
\lstick{x_{n}}& &\targ&\qw&\qw&\qw&\qw&\qw&\qw&\qw&\qw&\qw&\qw&\qw&\qw&\qw&\qw&\qw&\qw&\qw&\qw&\qw&\qw&\rstick{y_{n}}\\
&& T_{1,2,.,.n}^n&&& &  T_{1,2,..,n-1}^{n-1}&&&\  \ \ ...&&&T_{1,2,3,4}^{4}&&&T_{1,3,2}^3 && C_{1,3} && N_{1}&& C_{2,1}
}\\ \\ \ \ \ \ \ (a) &\ \ \ \ \ \ \ \ \ \ \ \ \ \ \ \ \ \ \ \ \ \ \ \ \ \ \ \ \ \ \ (b)
\end{tabular}
\end{footnotesize}
} %100 percent
\caption{The circuit representation for the decomposition of the
$R_{1,2,3,4,...,n}^n$ gate where: (a) the representation of the
gate, (b) the decomposition of the gate into it’s main
components gates.} \label{fig11}
\end{figure}

The $R^n$ gate is universal of $n$-bit gate, where it can be
extended according to the value of $n$. For $n\ge3 $, $R^n$
combines the action of the $n$ gates $N,C,T^3,T^4,..,T^{n-1},T^n$
as shown in Fig \ref{fig11}. There are $n!$ possible $R^n$ gates
which are sufficient to realize any $n$-bits circuit. The function
of the $n$! $R^n$ gates are shown in Eqn.\ref{eqn15}.\\
\begin{eqnarray}
\begin{array}{l}
R^n_{a_1,a_2,a_3,a_4,a_5,a_6,....,a_{n-1},a_n}: \\
\  y_{a_1}= x_{a_1}  \oplus x_{a3}  \oplus x_{a1}.x_ {a2} \oplus 1,\\
\ y_{a_2}= x_{a_2} \oplus x_{a_1}.x_{a_3} \oplus 1 , \\
\ y_{a_3}=x_{a_3} \oplus x_{a_1}, \\
\ y_{a_4}= x_{a_4} \oplus x_{a_1}.x_{a_2}.x_{a_3}, \\
\ y_{a_5}= x_{a_5} \oplus x_{a_1}.x_{a_2}.x_{a_3}.x_{a_4}, \\
\  y_{a_6}= x_{a_6} \oplus x_{a_1}.x_{a_2}.x_{a_3}.x_{a_4}.x_{a_5}, \\
\  ...\\
\  ...\\
\  ...\\
\ y_{a_{n-1}}= x_{a_{n-1}} \oplus x_{a_1}.x_{a_2}.x_{a_3}.x_{a_4}.x_{a_5}\ . \ . \ . \  x_{a_{n-2}}, \\
\  y_{a_n} \ \ \    = x_{a_n} \oplus x_{a_1}.x_{a_2}.x_{a_3}.x_{a_4}.x_{a_5}\ . \ . \ . \ x_{a_{n-1}}, \\
 \end{array}
\label{eqn15}
\end{eqnarray}
where $a_1$, $a_2$, $a_3$, .\ .\ .\  and $a_n$  $ \in \{1,2,3,...,n\}$  in any 
order. While the total number of possible gates for the general $T_n$  is  shown in Eqn.\ref{eqn15-1} \cite{citation09}.\\
\begin{eqnarray}
\begin{array}{l}
n\sum\limits_{r = 0}^{n - 1} {\left( {\begin{array}{*{20}c}
   {n - 1}  \\
   r  \\
\end{array}} \right)}, 
 \end{array}
\label{eqn15-1}
\end{eqnarray}
where $n$ is the number of bits and $r \ge 0$ is the number of controls per gate.\\
%
%%%%%%%%%%%%%%%%%%%%%%%%%%%%%%%%%%%%%%%%
\section{Experimental Results}
%%%%%%%%%%%%%%%%%%%%%%%%%%%%%%%%%%%%%%%%
This section compares the performance of the proposed two gate
libraries $R^3$ and $NR^3$, with the known libraries in
\cite{citation05}, \cite{citation06}, \cite{citation09},
\cite{citation21} and \cite{citation22}. It is found that the
permutation group of the $R^3$ generators is of size 40320, thus
the six $R^3$ generators are universal. There are 64 possible sub
libraries from the main $R^3$ gate library, 55 of them are
universal for the 3-bits reversible circuits. It is also found
that the permutation group of the $NR^3$ generators is of size
40320, thus the nine generators in $NR^3$ library are universal.
There are 512 possible sub libraries from the main library $NR^3$,
340 of them are universal. Table~1 compares the
utilization of the different universal libraries. It can be seen that 
the gate library $R^3$ gives the best utilization of 85.938\% 
and $NR^3$ gives a utilization of 66.406\%, which is better 
than the utilization of the libraries $NT$, $NP$, $NCT$ and $NCF$.\\

\begin{table}
\caption{\label{tbl1}Utilization of gates in universal sub Libs.}
\footnotesize\rm
\begin{tabular*}{\textwidth}{@{}l*{15}{@{\extracolsep{0pt plus12pt}}c}}
\br
\textbf{Lib} & \textbf{Lib Size} &
\textbf{\begin{tabular}[c]{@{}c@{}}Num of \\ sub Lib 
\end{tabular}} & \textbf{\begin{tabular}[c]{@{}c@{}}Num of universal \\ sub Lib\end{tabular}} &
\textbf{Utilization}\\
\noalign{\smallskip}\hline\noalign{\smallskip}
\textit{$NT$}   & 6                 & 64                       & 4                              & 6.250\%               \\ 
\textit{$NP$}   & 9                 & 512                      & 333                            & 65.039\%         \\ 
\textit{$NCT$}  & 12                & 4096                     & 1960                           & 47.852\%         \\ 
\textit{$NCF$}  & 12                & 4096                     & 2460                           & 60.059\%         \\ 
\textit{$NCP$}  & 15                & 32768                    & 26064                          & 79.541\%         \\ 
\textit{$NCTF$} & 15                & 32768                    & 23132                          & 70.593\%         \\ 
\textit{$NCPT$} & 18                & 262144                   & 217384                         & 82.925\%          \\ 
\textit{$NCPF$} & 18                & 262144                   & 220188                         & 83.995\%         \\ 
\textit{$G^3$}   & 6                 & 64                       & 51                             & 79.688\%            \\ \
\textit{$R^3$}   & 6                 & 64                       & 55                             &85.938\%             \\ 
\textit{$NR^3$}   & 9                 & 512                       & 340                            &66.406\%             \\ 
\br
\end{tabular*}
\end{table}

\begin{table}
\caption{\label{tbl2}Utilization of gates in the smallest universal sub libraries.}
\footnotesize\rm
\begin{tabular*}{\textwidth}{@{}l*{15}{@{\extracolsep{0pt plus12pt}}c}}
\br
\textbf{Lib}  & \textbf{\begin{tabular}[c]{@{}c@{}}Size of 
min uni-\\versal sub Lib\end{tabular}} &
\textbf{\begin{tabular}[c]{@{}c@{}}Num of sub Lib\\ with min
size\end{tabular}} & \textbf{\begin{tabular}[c]{@{}c@{}}Num of universal sub\\ Lib with min size\end{tabular}} &
\textbf{\begin{tabular}[c]{@{}c@{}}utilization\end{tabular}} \\
\noalign{\smallskip}\hline\noalign{\smallskip}
\textit{$NT$}   & 5	& 6	& 3	& 50\%\\
\textit{$NP$}   & 3	& 84	& 18	& 21.429\%\\  
\textit{$NCT$}  & 4	& 495	& 21	& 4.242\%\\ 
\textit{$NCF$}  & 4	& 495	& 60	& 12.121\%\\ 
\textit{$NCP$}  & 3	& 455	& 30	& 6.593\%\\ 
\textit{$NCTF$} & 4	& 1365	& 105	& 7.692\%\\ 
\textit{$NCPT$} & 3	& 816	& 36	& 4.412\%\\ 
\textit{$NCPF$} & 3	& 816	& 42	& 5.147\% \\ 
\textit{$G^3$}   & 2	& 15	& 9	& 60\% \\ 
\textit{$R^3$}   & 2	& 15	& 13	& 86.667\%\\ 
\textit{$NR^3$}   & 2	& 36	& 8	&22.222\%\\ 
\br
\end{tabular*}
\end{table}

\begin{table}
\caption{\label{tbl3}Minimum length of 3-bits reversible circuits using $NT$, $NP$, $NCT$, $NCF$, $NCP$, $NCTF$, $NCPT$, $NCPF$, $G^3$, $R^3$ and $NR^3$ libraries.}
\footnotesize\rm
\begin{tabular*}{\textwidth}{@{}l*{15}{@{\extracolsep{0pt plus12pt}}c}}
\br
 \textbf{\begin{tabular}[c]{@{}c@{}}Min \\ Len\end{tabular}}
& \textit{\textbf{NT}} & \textit{\textbf{NP}} &
\textit{\textbf{NCT}} & \textit{\textbf{NCF}} &
\textit{\textbf{NCP}} & \textit{\textbf{NCTF}}&
\textit{\textbf{NCPT}}& \textit{\textbf{NCPF}}&
 \textit{\textbf{G$^3$}} & \textit{\textbf{R$^3$}}& \textit{\textbf{NR$^3$}}\\
\noalign{\smallskip}\hline\noalign{\smallskip}
 0	& 1                    & 1                    & 1	
& 1                     & 1                     & 1
&1       &1 	&1	&1	&1\\  1
& 6                    & 9                    & 12
& 12                    & 15                    & 15
 &18     &18	&6	&6	&9\\ 2
& 24                   & 69                   & 102
& 101                   & 174                   & 143
&228     &248	&36	&33	&72\\  3
& 88                   & 502                  & 625
& 676                   & 1528                  & 1006
&1993    &2356 	&207	&180	&541\\  4
& 296                  & 3060                 & 2780
& 3413                  & 8968                  & 5021
&10503   &12797	&1097		&960		&3774\\  5
& 870                  & 13432                & 8921
& 11378                 & 23534                 & 15083
&23204 	&22794	&4946		&4686		&18027\\  6
& 2262                 & 21360                & 17049
& 17970                 & 6100                  & 17261
&4373		&2106		&13819	&14611	&17556\\  7
& 5097                 & 1887                 & 10253
& 6739                  & 0                     & 1790
&0 	&0	&14824	&15257	&340\\  8
& 9339                 & 0                    & 577
& 0                     & 0                     & 0
&0 	&0	&5208		&4555	 	&0\\  9
& 12237                & 0                    & 0
& 0                     & 0                     & 0
&0 	&0	&0	&31	&0\\  10
& 8363                 & 0                    & 0
& 0                     & 0                     & 0
&0 	&0	&0	&0	&0\\  11
& 1690                 & 0                    & 0
& 0                     & 0                     & 0
&0 	&0	&0	&0	&0\\  12
& 47                   & 0                    & 0
& 0                     & 0                     & 0
&0 	&0	&0	&0	&0\\  \textbf{Avg}
& \textbf{8.500}      & \textbf{5.516}      & \textbf{5.866}
& \textbf{5.649}       & \textbf{4.839}       & \textbf{5.330} 
& \textbf{4.730}	& \textbf{4.598}	& \textbf{6.403}
& \textbf{6.425}	& \textbf{5.325}
\\
\br
\end{tabular*}
\end{table}

The size of the minimum universal sub libraries from
 the main $R^3$ library and $NR^3$ library is two. 
For $R^3$, there are 13
universal sub libraries of size two, such as $\{
R_{1,3,2}^3, R_{2,1,3}^3\}$ , $\{ R_{1,3,2}^3,R_{2,3,1}^3\}$ and
$\{ R_{1,3,2}^3,R_{3,1,2}^3\}$, while for $NR^3$ there are 8
universal sub libraries of size two, such as $\{
R_{1,3,2}^3,R_{2,3,1}^3\}$, $\{ R_{1,3,2}^3,R_{3,1,2}^3\}$ and $\{
R_{3,1,2}^3,R_{3,2,1}^3\}$ .  Table~2 compares the
utilization of gates in the smallest universal sub libraries. The
utilization of the universal sub libraries with minimum size for
$R^3$ is 86.667\%, while for $NR^3$ is 22.222\%. It  shows that
$R^3$ gives the best utilization, while $NR^3$ is better than
$NP$, $NCT$,$NCF$, $NCP$, $NCTF$, $NCPT$ and $NCPF$. \\

Table~3 compares the minimum length for the 3-bits
circuits using different libraries. It shows that the average
minimum length for $R^3$ is 6.425, while for $NR^3$ it is 5.325
which is less than $NT$, $NP$, $NCT$, $NCF$, $NCTF$, $G^3$ and
$R^3$.\\

The optimization rules defined in \cite{citation21} has been
applied on the circuits built using $R$ , $NR$ and $NT$ Libraries, which can be summarized as follows:
first the $R$ gate is decomposed into it's five main components,
then the $T^3$ gate is decomposed into its five main components as
defined in \cite{citation21}. The adjacent gates are compared with
each other, if the two adjacent gates work on the same
input/output vectors, then they are merged to form one new gate,
such that: if gate $C^3_{i,j}$ is adjacent to gate $C^3_{i,j}$,
then the two gates are merged to give identity, otherwise if it is
adjacent to $C^3_{j,i}$ or $v_{j,i}$ or $u_{j,i}$, then these
gates are merged to form one gate. If $C^3_{i,j}$ is adjacent to
$v_{i,j}$, then they are merged to form $u_{i,j}$, otherwise if it
is adjacent to $u_{i,j}$, then they are merged to form $v_{i,j}$.
If $v_{i,j}$ is adjacent to $v_{i,j}$, then they are merged to
form $C_{i,j}$, or if it is adjacent to $u_{i,j}$, then these
two gates are merged to give identity, otherwise if it is adjacent
to $v_{j,i}$ or $u_{j,i}$, then they are merged to form one gate.
If $u_{i,j}$ is adjacent to $u_{i,j}$, then they are merged to
$C_{i,j}$, otherwise if it is adjacent to $u_{j,i}$ or $v_{j,i}$,
then these two gates are merged to form one gate, where $i,j,k$ 
$\in \{1,2,3\}$ in any order. Some adjacent gates can be swapped
with each other, if the target vector of one gate is not affecting
the controller vector of the other gate, as defined in
\cite{citation21}.\\

\begin{table}
\caption{\label{tbl4}Comparing the cost of the 3-bits circuits built using $R^3$, $NR^3$ and $NT$- gate libraries before and after optimization.}
\footnotesize\rm
\begin{tabular*}{\textwidth}{@{}l*{15}{@{\extracolsep{0pt plus12pt}}c}}
\br
\textbf{\begin{tabular}[c]{@{}c@{}}Min \\
Cost\end{tabular}} & \textbf{\begin{tabular}[c]{@{}c@{}}Num spc\\
in $R^3$ \\bfr optm.\end{tabular}} &
\textbf{\begin{tabular}[c]{@{}c@{}}Num spc\\
in $R^3$ \\aft optm.\end{tabular}}& \textbf{\begin{tabular}[c]{@{}c@{}}Num spc\\
in $NR^3$ \\bfr optm.\end{tabular}} &
\textbf{\begin{tabular}[c]{@{}c@{}}Num spc\\
in $NR^3$ \\aft optm. \end{tabular}} &
\textbf{\begin{tabular}[c]{@{}c@{}}Num spc\\
in $NT$ \\bfr optm.\end{tabular}}& \textbf{\begin{tabular}[c]{@{}c@{}}Num spc\\
in $NT$ \\aft optm.\end{tabular}}\\
\noalign{\smallskip}\hline\noalign{\smallskip}
\textit{0} & 1 & 1  & 7 & 7&7&7 \\ 
 \textit{1} & 0 & 0   & 0 &0&0&0   \\ 
\textit{2} & 0 & 0  & 0 & 0 &0&0  \\  
\textit{3} & 0 & 0& 0 & 0  &0&0 \\ 
 \textit{4} & 6& 6  & 192 & 192&0&0 \\  
\textit{5} & 0 & 0  & 0 & 0 &96&94  \\ 
\textit{6} & 0 & 0  & 0 & 0 &0&0  \\  
\textit{7} & 0 & 9& 0 & 851&0&0  \\  
\textit{8} & 33 & 24  & 3442 & 2591 &0&16 \\  
\textit{9} & 0 & 0  & 0 & 0 &0&340  \\ 
\textit{10} & 0 & 16  & 0 & 636 &648&288  \\  
\textit{11}& 0 & 68  & 0 & 6050 &0&32  \\  
\textit{12} & 180 & 96  &16040 & 9353  &0&179 \\  
\textit{13} & 0 & 28  & 0 & 39 &0&790 \\  
\textit{14} & 0 & 162  & 0 & 2829  &0&1487 \\
 \textit{15} & 0 & 394  & 0 & 7175  &2694& 324 \\ 
\textit{16} & 960 & 422  & 16676 & 6331   &0&574 \\ 
\textit{17} & 0 & 341  & 0 & 344 &0&2052 \\  
\textit{18}& 0 & 1121  & 0 & 1200  &0&3616 \\  
\textit{19} & 0 & 1919& 0 & 1278 &0&1462 \\  
\textit{20} & 4686 & 1798  & 3928& 1053  &7640&1041 \\  
\textit{21} & 0 & 1798  & 0 & 9  &0&3405 \\  
\textit{22} & 0 & 4218  & 0 & 14  &0&5357 \\
  \textit{23} & 0 & 5553  & 0 & 2  &0&2894 \\  
\textit{24} & 4059& 495  & 34 & 7 &0&1435 \\   
\textit{25}& 0 & 3097  & 0 & 0&12881&3191 \\   
\textit{26} & 0 &4934  & 0 & 0  &0&4369 \\   
\textit{27} & 0 & 4578  & 0 & 0&0&2436 \\   
\textit{28} & 15257 & 2410  & 0 & 0 &0&806  \\   
\textit{29} & 0 & 1407  & 0 & 0&0&1444  \\  
\textit{30} & 0 & 1273  & 0 & 0&11502&1482 \\  
\textit{31} & 0 & 524  & 0 & 0  &0&761 \\   
\textit{32} &4555 & 56  & 0 & 0 &0&125 \\   
\textit{33} & 0 & 4  & 0 &0  &0&126 \\   
\textit{34} & 0 & 4  & 0 & 0&0&109  \\
  \textit{35} & 0 & 0  & 0 & 0 &4489&60 \\  
\textit{36} & 31 & 0  & 0 & 0&0&6  \\   
\textit{37} & 0 & 0& 0 & 0&0&0 \\  
\textit{38} & 0 & 0  & 0 & 0 &0&0 \\
  \textit{39} & 0 & 0  & 0 & 0 &0&0 \\   
\textit{40} &0 & 0  & 0 & 0 &362& 0\\  
\textbf{Avg} & \textbf{25.701}& \textbf{23.954}  & \textbf{14.062} & \textbf{13.388}&\textbf{25.766}& \textbf{22.321} \\ 
\br
\end{tabular*}
\end{table}

The costs of all the 40320 specifications synthesized by the $R^3$,
$NR^3$ and $NT$ gate libraries are calculated and compared before
and after optimization in Table~4. For circuits built
using $R^3$-gate library, the maximum cost is 36 and the average
cost is 25.701 before optimization, while after optimization,
the maximum cost is 34 and the average cost is 23.954, giving
6.8\% of improvement. After adding the $N$ gate to the gate
library, the maximum cost has been reduced to 24, having an
average cost of 14.062 before optimization and 13.388 after
optimization, giving 44.1\% of improvement. For circuits built
using $NT$-gate library, the maximum cost is 40 and the average
cost is 25.766 before optimization, while after optimization
the maximum cost has been reduced to 36 and the average cost
became 22.321, giving 13.4\% of improvement. As shown in
Table \ref{tbl4}, the cost of the circuits built using $NR^3$-gate
library have improved by 40\%  over those built using $NT$-gate library.\\

\begin{table}
\caption{\label{tbl5}Optimal quantum cost of 3-bits reversible circuits using $NT$, $NCT$, $NCF$,  $NFT$, $NCTF$, $NCTPF$, $NR^3$ and $R^3$ libraries.}
\footnotesize\rm
\begin{tabular*}{\textwidth}{@{}c*{15}{@{\extracolsep{0pt plus12pt}}c}}
\br
\textbf{\begin{tabular}[c]{@{}c@{}}Min \\
Cost\end{tabular}} & \textbf{\begin{tabular}[c]{@{}c@{}}Num spc\\
in $NT$\end{tabular}} & \textbf{\begin{tabular}[c]{@{}c@{}}Num
spc\\  in $NCT$\end{tabular}} &
\textbf{\begin{tabular}[c]{@{}c@{}}Num spc\\  in $NCF$\end{tabular}} &
\textbf{\begin{tabular}[c]{@{}c@{}}Num spc\\ in $NFT$\end{tabular}}  &
\textbf{\begin{tabular}[c]{@{}c@{}}Num spc\\ in $NCTF$\end{tabular}} & 
\textbf{\begin{tabular}[c]{@{}c@{}}Num spc\\ in $NCTPF$\end{tabular}}&
\textbf{\begin{tabular}[c]{@{}c@{}}Num spc\\ in $NR^3$\end{tabular}}&
\textbf{\begin{tabular}[c]{@{}c@{}}Num spc\\ in $R^3$\end{tabular}} \\
\noalign{\smallskip}\hline\noalign{\smallskip}
0                & 7
& 7                                                            & 1
& 7                                                   &1  		&1	&5	&1\\   1                & 0
& 48                                                           & 9
& 48                                                    &9		&9	&0	&0\\   2                & 0
& 324                                                          &
51                                                           & 192
&51		&51		&0	&0\\
 3                & 0
& 607                                                          &
187                                                          & 408
&187  		&187		&0	&0\\
 4                & 0
& 601                                                          &
393                                                          & 480
&393		&405		&195	&6\\  5                & 94
& 1148                                                         &
474                                                          & 288
&477 		&609		&0	&0\\
 6                & 0
& 2462                                                         &
215                                                          & 592
&260  		&998		&0	&0\\
 7                & 0
& 3576                                                         &
17                                                           &
1962                              &338 	&2648		&851		&9\\  8                & 16
& 2710                                                         &
48                                                           &
3887   &1335		&4397		&2591		&24\\  9                &340
& 2855                                                         &
408                                                          &
2916   &3224		&2712		&0	&0\\  10               & 288
& 5601                                                         &
1919                                                         &
1299   &3686	&5994		&636		&16\\  11               & 32
& 6567                                                         &
3931                                                         &
3683 		&902		&10249	&6050		&68\\  12               & 179
& 3183                                                         &
2634                                                         &
7221 		&933		&1750		&9354		&96\\  13               & 790
& 2043                                                         &
462                                                          &
6059 		&4053		&3488		&396		&28\\  14               & 1487
& 3771                                                         & 5
& 1465                                                  &8690	&6640		&2829		&162\\  15               & 324
& 3496                                                         &
78                                                           &
3562  		&4903		&182		&7175		&394\\  16               & 574
& 1284                                                         &
1038                                                         &
4201       	&244		& 0		&6331		&422\\  17               & 2052
& 36                                                           &
6079                                                         &
2049         	&1094		& 0 	&344		&341\\  18               & 3616
& 0                                                            &
9571                                                         & 0
  &4346		& 0 	&1200		&1121\\
 19               &1462
& 0                                                            &
2036                                                         & 0
 &4724		& 0 	&1278		&1919\\
 20               & 1041
& 0                                                            &
12                                                           & 0
 &470		& 0 	&1053		&1798\\
 21               & 3405
& 0                                                            & 0
& 0    & 0  	& 0 	&9		&1798\\
 22               & 5357
& 0                                                            &
24                                                           & 0
  & 0  		& 0 	&14		&4218\\
 23               &2894
& 0                                                            &
732                                                          & 0
   & 0   	& 0 	&2		&5553\\
 24               & 1435
& 0                                                            &
5496                                                         & 0
 & 0  		& 0 	&7		&4059\\
 25               & 3191
& 0                                                            &
4482                                                         & 0
   & 0   	& 0 	&0		&3097\\
 26               & 4369
& 0                                                            &
18                                                           & 0
  & 0  		& 0	&0		&4934\\
 27               & 2436
& 0                                                            & 0
& 0            & 0  	& 0	&0 	&4578\\
 28               & 806
& 0                                                            & 0
& 0            & 0 	& 0 	&0		&2410\\
 29               & 1444
& 0                                                            & 0
& 0               & 0 	& 0 	&0		&1407\\
 30               & 1482
& 0                                                            & 0
& 0            & 0  	& 0 	&0		&1273\\
 31               & 761
& 0                                                            & 0
& 0         & 0		& 0 	&0	&524\\  32               & 125
& 0                                                            & 0
& 0           & 0 	& 0 	&0		&56\\  33               & 126
& 0                                                            & 0
& 0           & 0  	& 0 	&0	&4\\  34               & 109
& 0                                                            & 0
& 0             & 0 	& 0	&0 	&4\\  35               &60
& 0                                                            & 0
& 0        & 0		& 0 	&0	&0\\  \textbf{Avg} & \textbf{22.321}
& \textbf{10.348}                                         &
\textbf{17.468}                                         &
\textbf{11.770}   &\textbf{13.740}  &\textbf{10.520}      &\textbf{13.388}     &\textbf{23.954}                            \\ 
\br
\end{tabular*}
\end{table}

 Table \ref{tbl5} compares the minimum cost
for 3-bits circuits using different libraries. The average minimum
cost for the circuits built using $R^3$ is 23.954, while the average minimum cost for those built using
$NR^3$ is 13.388, which is less than those built using $NT$, $NCF$, $NCTF$ and
$R^3$.\\

%%%%%%%%%%%%% Conclution%%%%%%%%%%%%%%%%%%%%%%%
\section{Conclusion}
\label{sec:Conclusion}

We have presented new algorithms to synthesize exact universal
reversible gate library. Experimental results have been used to
compares the proposed universal gate library with the existing
approaches to reversible logic synthesis. We compares the minimum
cost for 3-bits circuits using different libraries. It is shown
that the cost of 40320 specifications synthesized by the $R^3$,
$NR^3$ and $NT$ gate library can be calculated and optimized.
Also, the average minimum cost for $R^3$ is 23.954, while the
average minimum cost for $NR^3$ is 13.388, which is less than
$NT$, $NCF$, $NCTF$ and $R^3$. For $n\ge3 $, $R^n$ combines the
action of the $n$ gates $N,C,T^3,T^4,..,T^{n-1},T^n$. There are
$n!$ possible $R^n$ gates which are sufficient to realize any
$n$-bits circuit. We have found that for $n$-bit circuits built
using $NR$-gate library, there are $n$! $R^n$ gates and $n$ $N$
gates. Finally, we have applied our optimal synthesis tool to
obtain much smaller circuits than previous methods.\\

%%%%%%%%%%%%%%%%%%%%%%%%%%%%%%%%%%%%%%%%%

%%%%%%%%%%%%%%%%%%%%%%%%%%%%

%%%%%%%%%%%%%%%%%%%%%%%%%%%%%%%%
\end{document}